\documentclass[a4paper,12pt,BCOR5mm,bibliography=totoc]{article}

\usepackage{caption}
\usepackage{subcaption}

\usepackage[english]{babel}
\usepackage{fancyhdr}
\usepackage{setspace}
\usepackage{graphics}
\usepackage{graphicx}
\usepackage{eso-pic}
\usepackage{amsmath}
\usepackage{amstext}
\usepackage{amsfonts}
\usepackage{amssymb}
\usepackage{hyperref}
\usepackage{slashed}
\usepackage{cite}
\usepackage{bbold}
\usepackage{multirow}
\usepackage{esdiff}
\usepackage[font=small,labelfont=small]{caption}
\usepackage{commath}
\usepackage{tikz-cd}
\usepackage{mathabx}
\usepackage{relsize}
\usepackage{simplewick} %/usr/local/texlive/2014/texmf-dist/tex/latex/simplewick/simplewick.sty

\usepackage[affil-it]{authblk} %affiliation
\renewcommand{\vec}[1]{\ensuremath{\boldsymbol{#1}}}

\newcommand{\iu}{\mathrm{i}}	%Symbol for the imaginary unit i
\newcommand{\ee}{\mathrm{e}}	%Symbol for the Euler number
\newcommand{\bra}[1]{\ensuremath{\left< #1\,\right|}}
\newcommand{\ket}[1]{\ensuremath{\left|\, #1\right>}}

\DeclareMathOperator{\Tr}{Tr}

\begin{document}

\vspace{0.01cm}
\begin{center}
{\Large\bf   Black Hole Type Quantum Computing  in Critical Bose-Einstein Systems} 

\end{center}

\vspace{0.1cm}

\begin{center}

{\bf Gia Dvali}$^{a,b,c}$ and {\bf Mischa Panchenko}$^{a}$\footnote{m.panchenko@campus.lmu.de} 

\vspace{.6truecm}

%\vspace{.2truecm}

{\em $^a$Arnold Sommerfeld Center for Theoretical Physics\\
Department f\"ur Physik, Ludwig-Maximilians-Universit\"at M\"unchen\\
Theresienstr.~37, 80333 M\"unchen, Germany}

%\vspace{.2truecm}

{\em $^b$Max-Planck-Institut f\"ur Physik\\
F\"ohringer Ring 6, 80805 M\"unchen, Germany}

{\em $^c$Center for Cosmology and Particle Physics\\
Department of Physics, New York University\\
4 Washington Place, New York, NY 10003, USA}

\end{center}

\vspace{0.5cm}

\begin{abstract}
\noindent  
 
{\small 
Recent ideas about understanding physics of black hole information-processing in terms of 
quantum criticality allow us to implement black hole mechanisms of quantum computing within  
critical Bose-Einstein systems.
The generic feature, uncovered both by analytic and numeric studies,  is the emergence at the critical point of gapless weakly-interacting modes, which 
act as qubits for information-storage at a very low energy cost.  
These modes can be effectively described in terms of either Bogoliubov or Goldstone degrees of freedom. 
The ground-state at the critical point is maximally entangled and far from being classical. 
We confirm this near-critical behavior by a new analytic method.  We compute growth of entanglement and 
show its consistency with black hole type behavior.   
On the other hand, in the over-critical regime 
the system develops a Lyapunov exponent and scrambles quantum information very fast. 
By, manipulating the system parameters externally, we can put it in and out of various regimes and in this way control the sequence of information storage and processing.  By using gapless Bogoliubov modes as control qubits we design some simple logic gates.

}

\end{abstract}

\thispagestyle{empty}
\clearpage

\section{Introduction}

 The recent advances \cite{cesargia, nico, scrambling, infogroup, gold} in understanding physics of black hole information processing in terms 
of the phenomenon of quantum criticality opens up 
 a possibility of studying and implementing its key mechanisms in ordinary Bose-Einstein systems, such 
 as for example, cold atomic gases.

Black holes are fascinating objects from the point of view of quantum information processing. 
The information storage capacity of a black hole of radius $R$ is measured by Bekenstein entropy \cite{Bent},
which scales as area in Planck units 
\begin{equation}
  N \, \sim \, {R^2 \over L_P^2} \, .
  \label{bek} 
\end{equation}
 The Planck area in terms of Planck and Newton constants is defined as $L_P^2 \equiv \hbar G_N$. 
Thus, the black hole entropy diverges in the classical limit, $\hbar \rightarrow 0$, if  $R$ and $G_N$ are kept finite.   In this limit black hole half life-time, $\tau_{BH} \, \sim \, N^{3/2}L_P$,  also becomes infinite. Hence, a classical black hole can store an infinite amount of information eternally. 

 What sort of microscopic physics is responsible for this type of behavior? 
According to \cite{cesargia}, the internal qubit degrees of freedom that store information in a black hole satisfy the following properties. 
 The energy gap in each qubit, is suppressed as some positive power of $1/N$ relative to  ``naive'' energy gap 
 that would be expected for an ordinary weakly-interacting quantum system of a similar size (for example, 
 for a gas of free cold bosons of mass $m$ in a box of size $R$, the gap between the levels would scale as $\sim \hbar^2/2mR^2$).  
  In the same time the interaction strength  of the qubit degree of freedom, with itself and the rest of the system, also must be suppressed by powers of $1/N$, in order to accommodate the fact that in $N\rightarrow \infty$ limit the 
 state of the qubit is an eigenstate of the Hamiltonian and the information stored in it is frozen eternally.  It was also suggested, that such a behavior 
 emerges from the following microscopic  physics.  The black hole can be described as  a bound-state 
 of  $N$ soft gravitons at the quantum critical point \cite{cesargia}. The quantum critical point amounts to the fact that the 
 graviton-graviton coupling $\alpha$  and their occupation number $N$ satisfy $\alpha N = 1$.   That is, the critical point  describes a quantum transition between the 
 free graviton Bose-gas and a collective bound-state.   It is this quantum criticality that gives rise to almost-gapless 
 collective excitations that play the role of the information qubits.  \footnote{Throughout the paper we use the term 
 ``qubit"  in generalized sense, although the near-gapless excitations living at the quantum critical point are in general multi-level quantum systems.  This allows for more possibilities of quantum information storage in these modes (e.g., using their coherent states \cite{gold}), but, for simplicity,  we can restrict ourselves to two basis-states per mode  that can be effectively separated from the rest.}

   We shall now abandon the gravitational aspect of this picture and take its main message, that the quantum criticality of
   Bose-gas is the key to the unusual features of black hole information processing.  From here it follows that similar properties must be exhibited by much simpler  systems that also can be placed at the quantum critical point,  such as the critical Bose-Einstein condensates of ordinary particles.  
    This expectation was confirmed  by  the previous studies, \cite{cesargia, nico, scrambling, gold}, where it was 
    shown that already the simplest prototype models,  indeed exhibit some of the key features of black hole informatics.

   One of the generic features is the appearance of nearly-gapless weakly-interacting qubits. 
   The important property is that the energy gap of qubits is controlled by closeness to criticality and can be made arbitrarily small.  In the same time, interaction of qubits is suppressed by $1/N$ and can be made arbitrarily 
   weak by choosing system with large $N$. 
    
   Another generic feature is the  high level of entanglement at the quantum critical point\cite{nico}. 
   It was also shown \cite{scrambling} that in the over-critical phase, where a Lyapunov exponent and chaos develops, the system 
   becomes a very fast scrambler of quantum information
   This result gives an explicit microscopic 
   justification to the conjecture of \cite{preskill} about the black hole's fast scrambling property.  It also 
   shows that this property is not limited to black holes, but is rather generic for critical Bose-Einstein systems with Lyapunov exponent and a chaotic behavior. \\

    All the above studies show that the effective theory of information storage near the quantum critical point is represented by 
    gapless and very weakly interacting qubit degrees of freedom. In these qubits one can store quantum information cheaply and for a macroscopically-long time. 
    
     In the present paper we take further steps in establishing the role of quantum criticality in black hole type  information processing.  In particular,  we focus on implementing this way of quantum computing in simple  Bose-Einstein systems.   
     
         We will be interested in the regime in which a system is very close to the quantum critical point. That is, in a situation  
       when a properly-normalized collective coupling, $\lambda \equiv \alpha N$, is extremely close to one. 
      As said above, this regime is universally characterized by appearance of almost gapless qubits.  For example, in Bogoluibov  
     approximation the energy gap scales as $\epsilon = \sqrt{1-\alpha N}$. However, the appearance of the nearly-gapless 
     mode is well-established also beyond this approximation, both by numeric \cite{nico,scrambling,gold}  as well as analytic methods \cite{gold}.   When approaching the critical point from the weak coupling ($\epsilon \rightarrow  0_+$)  the system is characterized by a slow evolution of qubits and a very sharp increase of  the entanglement  of the ground-state
     at $\lambda =1$.   Below we shall re-establish  this result analytically by using a new method. 
               
      In the overcritical regime, $\epsilon$ becomes imaginary.  There the system develops a Lyapunov exponent and becomes a fast scrambler.  \\
      
       By manipulating the parameters externally we can adjust proximity to the critical point and 
  take the system into different regimes.  In this way we can control the efficiency of information-storage 
  and processing. 
  
 We construct an explicit sequence of information storage and read-out in near-gapless Bogoliubov 
 modes of Bose-Einstein condensate  at the quantum criticality.   Using these modes as control qubits, we construct some quantum logic  gates. 
    We outline an explicit evolution sequence  that enables to store  information 
   in these qubits and later decode it.  \\

  We would like to note that the prototype Bose-Einstein systems, can be highly useful for exploring 
  other aspects of the black hole's quantum-critical portrait, such as, e.g.,  collapse and evaporation, \cite{NV}, \cite{thermal}.  In the present paper, however, we shall focus solely on the physics  of quantum information-processing.

\section{Cheap Qubits at Near Criticality} 

 We start with the system that  was used in \cite{cesargia} as the simplest prototype model for  exploring the  
connection between black hole information processing and quantum criticality. It represents a 
gas of cold bosons with a delta-function-type attractive interaction. 
The Hamiltonian in $d$-dimensions has the following form,     
   \begin{equation}
 {\mathcal H} \, = \, \int d^dx \, \psi^{+} \frac{- \hbar^2 \Delta}{ 2m} \psi \, - \, g \hbar  \ 
  \int d^dx \, \psi^{+}\psi^{+} \,\psi \psi \,, 
\label{Hnonderivative} 
\end{equation} 
where $\psi \, = \, \sum_{\vec{k}} \frac{1} {\sqrt{V}} {\rm e}^{i  {\vec{k} \over R} \vec{x}} \, a_k$, 
$V \, = \, (2\pi R)^d$ is the $d$-dimensional volume and $\vec{k}$ is the $d$-dimensional wave-number vector.  
$a_{\vec{k}}^\dagger, a_{\vec{k}}$ are creation and annihilation operators of bosons of momentum-number vector
${\vec{k}}$. These operators satisfy the usual algebra: $[a_{\vec{k}}, a_{\vec{k'}}^\dagger] = \delta_{\vec{k}\vec{k'}}$
and all other commutators zero. 
 The parameter $g$ controls the strength of the coupling. 
  The above system exhibits a quantum critical behavior.  In particular, in $d=1$, it is known to exhibit a quantum phase transition towards the bright soliton phase, studied in details in \cite{bright}.  
 
 It is useful to represent the Hamiltonian in the form,  
 $ {\mathcal H} \, \equiv \, {\hbar^2 \over 2R^2 m} \, H$.
  The quantity ${\hbar^2 \over 2R^2 m}$ is a convenient unit for energy-measurement, since 
  it gives us a clear idea about the energy-cost of a given process, 
  relative to the kinetic energy of a first non-zero momentum mode of a single free boson. 
  For this reason  all our further discussions will take place in the units ${\hbar^2 \over 2R^2 m} \, = \, 1$. 
   That is, we effectively switch to the Hamiltonian $H$.  

 Introducing a notation $\alpha \, \equiv \left ({g \over V R} \right) {2Rm \over \hbar} $ 
  and taking 
 $d=1$ this Hamiltonian takes the form,
 \begin{equation}
     H \, = \,  \sum_{k}  k^2 \, a_k^\dagger a_k  \, - \, {\alpha \over 4} \, \sum_{k_1+ k_2-k_3-k_4 = 0} 
   a_{k_1}^\dagger a^\dagger _{k_2}  a_{k_3}a_{k_4} \,.   
     \label{Hamilton}
 \end{equation}
 We restrict ourselves to three levels, $k=0,\pm1$.  Legitimacy of this approximation  for the regime of interest was firmly established  from the previous analysis of this system\cite{cesargia, nico, scrambling, gold,mischa}.  
  Let us define a triplet operator $a_{i} \equiv (a_{-1},a_0,a_1)$.
 Correspondingly,  we define the number operators 
 $n_{i} \equiv a_{i}^\dagger a_{i}$. Since the particle number is conserved we can restrict ourselves to an $N$-particle sector.
  
   Then the Hamiltonian in terms of the triplet $a_{i}$ can be written as \cite{gold},
 \begin{equation}
     H \, = \,  \sum_{i=-1,1}\,  n_{i} \, -  \, {\alpha \over 2} \, (a_1^\dagger a^\dagger _{-1}  a_0a_0 \, + \, 
a_0^\dagger a^\dagger _0  a_1a_{-1})  \, + \, {\alpha \over 4} \sum_{i=-1}^1 \, n_{i}^2 \, + \, 
 H(n) \,.   
     \label{Hamilton1}
 \end{equation}
 where,  $H(n)$ is the part of the Hamiltonian that depends on the total number operator $n \equiv n_{-1} +n_0+n_{1}$ and has the following form,  
  \begin{equation}
    H(n) \, = \,  - \,  {\alpha \over 2} \, n^2 \, + \,  {\alpha \over 4} \, n \, + \, \mu \,  (n \, - \, N) \, ,
     \label{HamiltonHn}
 \end{equation}
 where $\mu$ is a Lagrange multiplier. 
  For $\alpha N \ll 1$ the minimum of the Hamiltonian with a very good approximation is achieved 
at the state $n_{i}  \, = \, (0,N,0)$.  Expanding around this state the unsuppressed part of the Hamiltonian 
can be expressed as the mass matrix that mixes   the operator doublets $(a_1^\dagger ,  a_{-1})$ and 
$(a_1 ,  a_{-1}^\dagger)$,
%effective mass matrix 
%for $a_1$ and $a_{-1}$ modes, 
 \begin{equation}
  \label{eq:massmat}
   \begin{pmatrix}
     1\,- \, {\alpha \over 2}N \,, & -\, {\alpha \over 2}N  \\
     - \, {\alpha \over 2}N \,, & 1\, - \,{\alpha \over 2} N 
\end{pmatrix} \, .
\end{equation}
 We shall now  diagonalize this bilinear hamiltonian via a Bogoliubov canonical transformation, 
\begin{equation}
b_{\pm 1} = u a_{\pm 1} - v a_{\mp 1}^\dagger\,,
\end{equation}
where, 
\begin{equation}
u = \frac{1 + \epsilon}{2\sqrt{\epsilon}}\,,\hspace{2em}v = \frac{1 - \epsilon}{2\sqrt{\epsilon}}\,,  
\end{equation} 
with
 \begin{equation}
\epsilon \, = \, \sqrt{1-\alpha N} \,.
\label{eq:bogen}
\end{equation}

We thus arrive at the following effective low energy Hamiltonian for Bogoliubov modes:
 \begin{equation}
 H_b \,  = \,  \epsilon (b^\dagger_{-1} b_{-1} +  b^\dagger_{+1} b_{+1} )  \, + \, {1 \over N \epsilon^2} {\mathcal O} (b^4) \,, 
 \label{effective} 
 \end{equation}
 where ${\mathcal O} (b^4)$ stands for momentum-conserving quartic interactions among $b_{\pm1}$ and $b^\dagger_{\pm1}$ operators. 
  
  The proximity of the system to the quantum critical point can be measured by the parameter $\epsilon$.
  This parameter can be tuned to be arbitrarily small by choosing $\alpha N$ sufficiently close to $1$. 
   Moreover, for 
  any given value of  $\epsilon>0$ we can always take large-enough $N$ so that the interaction term in (\ref{effective}) 
 can be neglected for low-enough $b$-levels.   We shall refer to this choice of system parameters $\epsilon$ and $N$,
 as a double-scaling regime of the Bogoliubov domain.    
 
  The validity range of this regime will be quantified below. 
 Before doing this, we would like to pause and reflect on the fundamental importance of the   Hamiltonian $H_b$ for quantum information storage and processing.   This connection was already much appreciated in \cite{cesargia,nico,scrambling,gold} and is also central to the present work.  Namely, in the double-scaling regime $H_b$ becomes a free Hamiltonian 
 of $b$-modes with energy gap $\epsilon$.  Thus, the states of small-enough occupation number of $b$-particles,  
 $|n\rangle_b$, are approximate
 energy eigenstates, $H_b |n\rangle_b \, = \, E_n |n\rangle_b$  of eigenvalues  $E_n \, = \, n\epsilon$.
 Here and below $n$ denotes an eigenvalue of the  $b$-number operator, $b^+b$.   As  said above, for each 
 value of $\epsilon$ this approximation 
 can  be made arbitrarily accurate by taking $N$ to be sufficiently large. The precise scaling will be quantified below. 
 \\

  Thus, using $b$-modes as qubits, the 
 energy-cost of information-storage can be made {\it arbitrarily cheap} as compared to typical energy 
 gap in the ordinary non-critical systems. Secondly, the information stored in $b$-modes stays intact for a macroscopically-long time, controlled by the powers of $N$.

     \section{Regimes}  
   Let us now quantify the parameter-choices for various regimes. We start with the double-scaling regime of Bogoliubov domain
 in which, as described above,  Bogoliubov effective theory can be made applicable by taking 
 both $\epsilon$ and $1/N$ sufficiently small.  We shall identify the required relation between $\epsilon$ and $N$. 
    
    In order to achieve this,  let us notice that  for any fixed values of $N$ and $\epsilon$ there exists 
   a maximal occupation number $n_{max}$,  above which the Bogoliubov treatment breaks down.    
   Obviously this value of  $n$ also measures the number of states  that fall within Bogoliubov   
  treatment.   
  This number can be easily estimated from the following argument. 
  
  First, we shall require that the energy of the state must be within the gap  
 $\Delta H_b \sim 1$. From the form (\ref{effective}) it is clear that this condition translates as the bound on occupation number of $b$-modes,   
  \begin{equation}
  n \, < \, \epsilon^{-1} \, .     
  \label{condition1}
  \end{equation}
 Secondly,  the validity of Bogoluibov approximation imposes a condition that the interaction term among $b$-modes must be sub-dominant with respect to the first term 
 in  (\ref{effective}).  This condition translates as a second bound on $n$, 
 \begin{equation}
  n \, < \, \epsilon^3N \, .    
  \label{condition2}
  \end{equation}
   The maximal value $n_{max}$  that satisfies both conditions is bounded from above as   
 \begin{equation}
  % \epsilon_{min}\, \sim N^{-1/4},~~{\rm and~correspondingly} \,, ~~ 
   n_{max} \, < \, N^{-1/4} \, .    
  \label{conditions}
  \end{equation} 
 Thus, for any $N$ we can adjust $\epsilon \, \sim \, N^{-1/4} $ in such a way that there are $\sim N^{1/4}$ approximate energy eigenstates within the $\Delta H_b \sim 1$ energy gap.   The storage of information within the lowest lying states is extremely cheap and the evolution time is long. 
 
    We can determine the absolute bound on the domain of Bogoliubov treatment,  by requiring that there exist  at least one number-eigenstate within its validity.  That is, we take $n \sim 1$.  Then from 
 (\ref{condition2}) we get the absolute lower  bound on $\epsilon$, 
 \begin{equation}
  % \epsilon_{min}\, \sim N^{-1/4},~~{\rm and~correspondingly} \,, ~~ 
   \epsilon \, > \, N^{-1/3} \, .    
  \label{conditions}
  \end{equation} 
  Thus, by taking $N$ sufficiently large we can make the energy gap arbitrarily-small and controllable 
  within the effective theory given by (\ref{effective}). \\
  
  The bound (\ref{conditions}) marks the domain of applicability of the Bogoliubov treatment. 
  However,  near-gapless states continue to exist even beyond this domain.  
  In fact, the alternative methods \cite{scrambling, gold}  as well as the 
  new one that will be discussed below show that 
   density of such states in the vicinity of critical point remains huge. 
  However,  they cannot be analyzed within the Bogoliubov effective theory given by (\ref{effective}). 
 In this domain 
  $\epsilon$ is still real, but it is below the Bogoliubov bound,  $\epsilon \,< \, N^{-1/3}$. 
  We shall refer to this domain as to trans-Bogoliubov. 
   In this domain the Bogoliubov 
  approximation is no longer valid and we have to employ a different method. 
   In \cite{nico, scrambling} this was achieved by  exact numerical diagonalization and in \cite{gold} by developing an effective 
  Nambu-Goldstone description of modes.  Below we shall use a different method for treating this domain. \\
     
  Finally, there is a third domain, $\alpha N > 1$, in which  $\epsilon$ becomes imaginary.   In this domain, the state in which only $n_0$-level is macroscopically-occupied becomes unstable and system relaxes to a new ground-state of a bright soliton.   As shown in \cite{scrambling}, 
 during this instability the system is a very efficient scrambler of 
  information due to its Lyapunov exponent and a very high density of states near the critical point. \\
  
   To summarize, we identify three domains of interest in the vicinity of quantum criticality.  The two 
   belong to the sub-critical regime ($\alpha N <1$, i.e., $\epsilon > 0$).  
   The first of these two is  the double-scaling Bogoliubov domain, in which $\epsilon$ is real and  satisfies the bound  (\ref{conditions}). 
    The second is  trans-Bogoliubov domain, in which $0 < \epsilon  <  N^{-1/3}$.
   Both of these domains are characterized by a very low energy gap as well as by slow evolution of low-lying states. In particular, generation of entanglement takes macroscopic time, enhanced by powers of $N$. 
       
    Finally, there is an over-critical
    fast-scrambling domain with imaginary $\epsilon$,  in which the system becomes unstable.  Unlike the other two this domain is characterized by an extremely fast growth of the entanglement and scrambling of information. \\

   Having identified the different domains, we can manipulate the system putting it in and out of different 
   regimes in accordance with the information-processing task.  This can be done, because by manipulating  
  characteristics of the Bose-Einstein system we can change criticality parameters, such as $\epsilon$ (or $\alpha N$). 
  
  For example, consider the following sequence of information storage and processing.  
  First, keeping system within Bogoliubov domain, we can dial $b$-modes by coupling the system 
  temporarily to some external influence.  This dial-up encodes the initial message in form of the occupation numbers of $b$-modes. 
 Then, after switching-off the external influence, information remains stored 
  in $b$-modes for a long time due to very slow evolution.  Next, by switching on the coupling with some  external qubits, we can process the stored information by using $b$-modes as control qubits.  We give  examples of designing some simple logic gates in the next section.   
 
  Alternatively, putting the system in over-critical domain, we can scramble the information and transform the original message in a highly-entangled state.  In this regime $\epsilon$ becomes imaginary.  The system acquires a Lyapunov exponent and scrambling 
 time becomes logarithmic in $N$.  
     In principle, by manipulating the parameter $\epsilon$ externally, one can control the level of entanglement.

 \section{Qubit Evolution}
 As an example of using Bogoliubov modes as qubits near the critical point, consider the following 
system.  Let the Bogoliubov mode operator be $b$. Near the critical point in the double scaling limit  the  Hamiltonian  (\ref{effective})  can be approximated as a free Hamiltonian of $b$-modes. Let us
consider a single lowest lying $b$-mode, 
 \begin{equation}
     H_b \, = \, \epsilon \,  b^\dagger b  \,. 
     \label{Hamilton-b}
 \end{equation} 
We assume that  $ N^{-1/4} \, > \, \epsilon \, > \,  N^{-1/3}$ in order to be 
within the validity domain of the Bogoliubov approximation and ignore $1/N$-suppressed interaction terms.   

  Let us couple $b$ to a probe external quibit  $c$ with creation annihilation algebra $[c, c^\dagger]=1$. We choose the Hamiltonian in the following form,  
 \begin{equation}
     H \, = \, \epsilon \,  b^\dagger b  \,  +   b^\dagger b (\nu c + \nu^* c^\dagger) \, + \, 2\delta c^\dagger c \, .
     \label{Hamilton-b1}
 \end{equation} 
   Here $\delta$ is a real parameter and $\nu$ is complex, which for simplicity we choose to be real as well.  
  For non-zero coupling $\nu$ the two degrees of freedom can control the time-evolution of each other. In particular 
  $c$ can be used for encoding information in a state of $b$-modes, as well as for reading-out this information.     
 
 \subsection{$b$-Mode as Control Qubit} 
 Let us first consider a situation in which $b$ acts as a control qubit for $c$.  For this purpose it is enough to 
 take $c$ as a two-level system described by the states $|0\rangle_c$ and $|1\rangle_c$.   
    
For $|\nu| \ll \sqrt{\epsilon \, \delta}$, the ground-states of the two systems are uncorrelated and are given by 
a tensor product of two free-oscillator ground-states, $|0\rangle_b\otimes |0\rangle_c$.  However, for 
 $|\nu| \gg \sqrt{\epsilon \, \delta}$  the two systems become strongly correlated.

   In particular, in the regime $\epsilon \sim \nu \gg \delta$, the ground-state of the qubit  $c$ is determined by the state of $b$.  Thus, $b$ acts as a control qubit for $c$. \\

 It is easy to compute the time evolution as given by (\ref{Hamilton-b1}). If we label the basis states as
$$\left( \begin{matrix}
\alpha \\ 
\beta
\end{matrix} \right)_m
:=\alpha\ket{m}_b\ket{0}_c+\beta\ket{m}_b\ket{1}_c \, ,$$ 
we find
 \begin{equation}
 H \left(\begin{matrix}
 \alpha\\ \beta
 \end{matrix} \right)_m = A_m\cdot \left(\begin{matrix}
 \alpha\\ \beta
 \end{matrix} \right)_m,
 \end{equation}
    with $$A_m=\left(\begin{matrix}
    \varepsilon m && \nu m\\
    \nu m && \varepsilon m +2\delta
    \end{matrix}\right).
$$  
The Hamiltonian becomes \textit{block diagonal} in this representation. We can now evolve any qubit simply by exponentiating $A_m$. For time independent $\nu$ we find the time evolution of states to be
 \begin{align}\label{timevqub}
\left(\begin{matrix}
 \alpha\\ \beta
 \end{matrix} \right)_m(t) &= \ee^{-\iu t H}\left(\begin{matrix}
 \alpha\\ \beta
 \end{matrix} \right)_m = \nonumber \\
& \frac{\ee^{-\iu t (\delta +m\varepsilon )}}{\omega_m} 
 \left( \begin{matrix}
\alpha \, \omega_m  \cos (t \omega_m )+\iu \sin (t \omega_m ) (\alpha  \delta -\beta  m\nu) \\
 \beta \, \omega_m  \cos (t \omega_m )-\iu \sin (t \omega_m ) (\beta  \delta +\alpha m\nu)
\end{matrix}
\right)_m
\end{align}
where $\omega_m:=\sqrt{(m\,\nu)^2+\delta^2}$. In particular, for $m=0$:
\begin{equation}
 \left(\begin{matrix}
 \alpha\\ \beta
 \end{matrix} \right)_0(t) =\left(\begin{matrix}
 \alpha\\ e^{- i t 2 \delta}\beta
 \end{matrix} \right)_0.
\end{equation}

For time dependent $\nu=\nu(t)$ one merely has to substitute $t\nu$ with $\int \nu(t) dt$.

 We see that the Hamiltonian acts as a generalized controlled gate, in the sense that there are two different time scales for qubit state with $0$ occupation number of  $b$-particles $(m=0)$ and with one or more $b$-particles $(m>0)$. For the $m=0$ state the time scale is $\frac{1}{\delta}$ whereas for $m>0$, after the adiabatic turning on of $\nu(t)$ until the final value $\nu_{end}\gg\delta$, the time scale is given by $\frac{1}{\sqrt{(m\,\nu_{end})^2+\delta^2}}\approx\frac{1}{m\,\nu_{end}}\ll \frac{1}{\delta}$. For $\nu_{end}\gg\delta$ the $m=0$ state can be viewed as stationary.\\

   For example, restricting to two levels $m=0,1$ in the limit $\delta=0$ and for $t\nu = \pi/2$ this system acts as a standard  CNOT gate, because it acts trivially on the state with $m=0$, whereas for $m=1$ we have (up to a phase factor), 
   $$\left( \begin{matrix}
\alpha \\ 
\beta
\end{matrix} \right)_1
\rightarrow  \left( \begin{matrix}
\beta \\ 
\alpha
\end{matrix} \right)_1 \,.$$

     \subsection{Encoding Information in $b$-Mode via Time Evolution with Coherent States} 
      
  For encoding the initial information in $b$-mode an external mode with similar coupling can be used.
  We keep the notation $c$ for this mode. 
  We shall consider the case when the state of $b$-mode is controlled by the coherent state of $c$.
   Preparing the state of $c$ in a coherent state $|\gamma\rangle_c$ and switching on coupling 
   $\nu(t)$, we can evolve the $b$-mode to a particular state controlled by the coherent state parameter 
   $\gamma$.  
    We take the Hamiltonian to be (\ref{Hamilton-b1}) again, however, now the operator $c$ is no longer restricted to two levels, but is a fully-fledged harmonic oscillator. We want to study the time evolution of $b$-particle states tensored with coherent states of $c$. Again the problem can be treated analytically in simple terms. We now compute
  \begin{equation}\label{coherenttimeev}
  \ee^{-\iu\, H\, t} \ket{m}_b\otimes \ket{\gamma}_c \,,
  \end{equation}
     where $\ket{\gamma}_c$ is a normalized coherent state of $c$, i.e., $c\ket{\gamma}_c=\gamma \ket{\gamma}_c$. We get
 \begin{equation}
 (\ref{coherenttimeev})= \ee^{-\iu \, \epsilon \, m t}\ket{m}_b \otimes \ee^{-\iu t( m\nu(c+c^{\dagger})+2\delta\, c^{\dagger}c)}\ket{\gamma}_c.
 \end{equation}
We now perform the following trick.  Let us define $d:=c+\frac{m\nu}{2\delta}$. Note that coherent states for $c$ are also coherent states for $d$, namely
\begin{equation}
d\ket{\gamma}_c=\left(\gamma+\frac{m\nu}{2\delta}\right)\ket{\gamma}_c \Longleftrightarrow 
\ket{\gamma}_c=\ket{\gamma+\frac{m\nu}{2\delta}}_d.
\end{equation} 
Then 
\begin{equation}
 2\delta c^{\dagger}c+m\nu(c+c^{\dagger})=2\delta \, d^{\dagger}d-\frac{(m\nu)^2}{2\delta}
\end{equation}
and we find
\begin{equation}
  \ee^{-\iu t( m\nu(c+c^{\dagger})+2\delta\, c^{\dagger}c)}\ket{\gamma}_c=\ee^{\,\iu t \frac{(m\nu)^2}{2\delta}}\ee^{-\iu t 2\delta \, d^{\dagger}d }\ket{\gamma+\frac{m\nu}{2\delta}}_d.
\end{equation}
The last equation is simply the time evolution of a coherent state with frequency $2\delta$, so that in total we find:
\begin{equation}
 \ee^{-\iu\, H\, t} \ket{m}_b\otimes \ket{\gamma}_c= \ee^{-\iu t(\epsilon m -\frac{(m\nu)^2}{2\delta})}\ket{m}_b \otimes \ket{\ee^{-\iu t 2\delta}\left(\gamma+\frac{m\nu}{2\delta}\right)-\frac{m\nu}{2\delta}}_c.
\end{equation}
The time evolution preserves coherence and the speed of the evolution depends on $\abs{\gamma+\frac{m\nu}{2\delta}}$. Incidentally we observe that states $\gamma=-\frac{m\nu}{2\delta}$ do not evolve, i.e., are energy eigenstates with energy $(\epsilon m -\frac{(m\nu)^2}{2\delta})$. 
\footnote{Obviously, the Hamiltonian is unbounded from below for time-independent $\nu$}. 
\\

We note that we can dial $b$-states, e.g., by first preparing a particular coherent state in $c$, say $(\sum \beta_j\ket{m_j}_b)\otimes \ket{-\frac{m_0\nu}{2\delta}}_c$, where $\beta_j$ are some parameters describing the state of $b$, 
and then after some time probing it against itself. The matrix element
\begin{equation}
 \left(_c\bra{-\frac{m_0\nu}{2\delta}}\otimes\, _b\bra{X}\right) U_t \left( \sum \beta_j\ket{m_j}_b\otimes  \ket{-\frac{m_0\nu}{2\delta}}_c\right)
\end{equation}
is dominated by the contribution from $\ket{m_0}_b$ for any $\ket{X}_b$ (as long as it has a non-zero overlap with $\ket{m_0}_b$) since 
\begin{equation}
 \bra{-\frac{m_0\nu}{2\delta}} \left. \ee^{-\iu t 2\delta}\frac{(m_j-m_0)\nu} {2\delta}-\frac{m_j\nu}{2\delta} \right>_c =\ee^{-\frac{\nu^2}{2\delta^2}(m_j-m_0)^2\sin^2(\delta t)-\iu(...)}
\end{equation}
is exponentially suppressed for $\nu\gg\delta$ for any $m_j\neq m_0$ when averaged over time.
Here,  the phase factor in the exponent is  $(...)= {\nu^2\over 4\delta^2} m_0(m_0-m_j) {\rm sin} (2t\delta)$.

Thus, if one has a pointer which reacts to classical configurations of $c$, by preparing  and measuring states with such a pointer one can collapse randomly-generated states to fixed $b$ mode states - i.e., one can prepare certain $b$-mode states  by performing measurements on $c$ only.

\section{Beyond the Bogoliubov Domain}
We now wish to study evolution of the system near the critical point in the domain $0< \epsilon < N^{-1/3}$,  which is  
beyond the validity of the Bogoliubov double-scaling domain.  In \cite{nico, scrambling,gold} this regime was studied numerically.  In \cite{gold}  a new method was developed in which the quantum phase transition 
was mapped onto a Goldstone phenomenon in a sigma-model that describes a spontaneous breaking of the  
global  $U(3)$-symmetry of Hamiltonian (\ref{HamiltonHn}), under which the triplet 
of operators $(a_{-1}, a_{0}, a_{1})$  transforms in the fundamental representation. 
In the state $(0, N, 0)$ this symmetry is spontaneously broken down to $U(2)$.  The quantum phase transition 
corresponds to a change of symmetry breaking pattern to a smaller symmetry group achieved on the ground-state $(x, N-2x, x)$.  As shown in \cite{gold} the transition to this new ground-state with $x\neq 0$ takes place beyond 
$\alpha N =1$ point.  
 This mapping allows to relate the gapless mode, appearing at the quantum critical point, to a Nambu-Goldstone  
mode of the sigma-model that becomes massless at the same transition point.  The effective Hamiltonian  of the gapless mode 
up to $1/N$-corrections is $H_{gold} \, = \, {n_{gold}^2 \over N}$, where $n_{gold}$ is the occupation number of the gapless mode.  The consistency of this effective description was also checked by numerical study of 
various characteristics, such as,  derivation of the ground-state, calculation of the energy for different occupation numbers of Goldstone mode and exact time-evolution of states \cite{gold}.

Thus, the effective Goldstone approach shows that the existence of nearly-gapless weakly-coupled excitations near the critical point persists both in trans-Bogoliubov domain, $0 \,  < \, \epsilon \, < N^{-1/3}$, 
as well as in the over-critical one, as long as  $|\epsilon|$ (equivalently $x$) is small.

 Below, we shall complete the study by applying a new method, 
which allows to treat the system analytically beyond the Bogoliubov approximation in sub-critical domain and compute 
certain important characteristics, such as the growth of entropy at the critical point, explicitly. 
   In this paper  we shall only outline the technique that is essential for our analysis. The detailed description of the original method will be given in \cite{mischa}.      

\subsection{The Setup} 

 As discussed above, after we restrict ourselves to an $N$-particle sector with total momentum zero and cut off the modes with momenta 
$|k| > 1$, we arrive at the Hamiltonian (\ref{Hamilton1}). 
The Hilbert space becomes finite dimensional. Its basis can be written as $$\ket{n}:=\ket{n_{-1}=n,n_0=N-2n,n_{1}=n},$$
 where $n$ goes from zero to $\frac{N}{2}$. The three mode Hamiltonian in this basis is a \small{$(\frac{N}{2}+1)\times (\frac{N}{2}+1)$} dimensional tridiagonal matrix of the following form:
\begin{equation}\label{matrix}
H= \begin{pmatrix}
d_0   & h_{1} & 0  & \dots \\
h_1 &  d_1   & h_{2}  & \ddots \\
0& h_2 & \ddots & \ddots \\
\vdots & \ddots &\ddots &\ddots
\end{pmatrix}
-c\,N,
\end{equation}
where the entries are 
\begin{equation}
 d_n=(2-\lambda)n+\frac{3\lambda}{2}\frac{n^2}{N} \quad,\quad h_n=-\frac{\lambda}{2}n+\frac{\lambda}{N}n^2 \, .
\end{equation}
Here $\lambda \, \equiv \, \alpha N$ is the same parameter as defined in the introduction and $c$ is a positive constant \footnote{We have approximated $\sqrt{(N-2n)(N-2n-1)}\approx (N-2n)$.}.

  The spectrum and the eigenstates of $H$ were computed analytically by means of a novel diagonalization technique. We will not explain this technique  here, but rather state the dictionary provided in \cite{mischa} shortly and refer the reader to the original paper for more details. This dictionary is
%\begin{equation}
\begin{align*}
\ket{n} &\longleftrightarrow b_n(x):=\ee^{-\frac{x}{2}}L_n(x) \\
H &\longleftrightarrow \frac{\lambda}{2}(x-1)+2(\lambda-1)\left(x\,\partial_x^2+\partial_x+\frac{2-x}{4}\right)+\frac{1}{N}H_{int}.
\end{align*}
%\label{dict}
%\end{equation}

\subsection{The critical entropy growth}
\subsubsection{Entanglement of the condensate state}
At the critical coupling $\lambda=1$ the unsuppressed part of $H$ corresponds to a multiplication operator. The suppressed interaction part $\frac{1}{N}H_{int}$ is extremely important for the computation of the eigenvalues and eigenstates of $H$ - indeed it controls the depletion and regulates the phase transition. However, for states in which only low $\ket{n}$ are occupied the interaction part does not contribute to the time evolution for long times for large $N$ \footnote{This is justified both by our previous discussion as well as by the existing studies \cite{cesargia, gold}  that show that evolution takes macroscopic time in $N$.}. Therefore we can use the multiplication operator for computing the time evolution of those states. This greatly simplifies the problem.\\

We wish to compute the time evolution of the condensate state $\ket{0}$, in other words, the coefficient
\begin{equation}\label{an}
\kappa_n:=\bra{n}\ee^{-\iu \, H t}\ket{0}.
\end{equation}
From $\kappa_n$  we can calculate the 1-particle density matrix and then the von-Neumann entropy of the time-evolved condensate can be calculated as function of time. The density matrix $\rho$ in this context is a $3\times 3$ diagonal matrix where the entries are \cite{nico}
\begin{align}
\begin{split}
\rho_{00}=\frac{1}{N}\sum\limits_{n=0}^{\frac{N}{2}}\abs{\kappa_n}^2(N-2n)  \\
\rho_{11}=\rho_{-1\,-1}=\frac{1}{N}\sum\limits_{n=0}^{\frac{N}{2}}\abs{\kappa_n}^2 n \, .
\end{split}
\end{align}
 \\

Now we use the fact that  for a long time-scale (proportional to some positive power of $N$, to be determined below) and for small enough $n$, we can use the multiplication operator to compute $\kappa_n$.
It turns out that ``small enough" $n$ are the dominant contribution to the density matrix, so we are in a very comfortable position. From the dictionary above we hence obtain:
\begin{equation}\label{coefficients}
\kappa_n\approx\int_0^{\infty } \ee^{-x} \ee^{\frac{-\iu}{2} t (x-1)} L_n(x) \, dx= \ee^{\frac{\iu}{2} t} \frac{2(-\iu t)^n}{(2+\iu t)^{n+1}},
\end{equation}
where the approximation is very good for small $n$ over long time scales. From here we compute:
\begin{align}\label{rho}
\begin{split}
\rho_{00}=\frac{t^2 }{2 N} \left(\left(\frac{t^2}{t^2+4}\right)^{\frac{N}{2}}-1\right)+1 \\
\rho_{11}=\frac{t^2}{4N}\left(1-\left(\frac{t^2}{t^2+4}\right)^{\frac{N}{2}}\frac{2 N+t^2+4}{t^2+4}\right) \, .
\end{split}
\end{align}
The von Neumann entropy is given by $-\Tr(\rho\log\rho)$, i.e., by:
\begin{equation}
 S=-\left(\rho_{00}\log(\rho_{00})+2\rho_{11}\log(\rho_{11})\right) \, .
\end{equation}
For small times and large $N$ we find 
$$\rho_{00}\approx -\frac{t^2 }{2 N}+1 \quad,\quad \rho_{11}\approx\frac{t^2}{4N},
$$
and so for early times the entropy grows as
\begin{equation}
 S\sim-\log\left(\frac{t}{\sqrt{N}}\right) \frac{t^2}{N}.
\end{equation}

 Thus the characteristic time for the entropy growth is macroscopic and scales as $t_{ent} \sim \sqrt{N}$.  
 This confirms the earlier findings \cite{nico, scrambling, gold}  indicating that the evolution of lowest 
 lying states remains slow, even beyond the Bogoliubov regime, as long as the system is subcritical.
 It is also in agreement with the derivation of entanglement near the critical point given in
 \cite{nico}. \\

We compare the analytic expression for $S$ with the one obtained numerically using the matrix (\ref{matrix}) for $N=250$, see the image below. We see that even for such a small $N$ the analytical expression for $S$ is valid for long times, i.e., the entropy indeed grows until its maximal value with the relevant time-scale being $\sqrt{N}$.

\begin{figure}[h]
\includegraphics[scale=0.8]{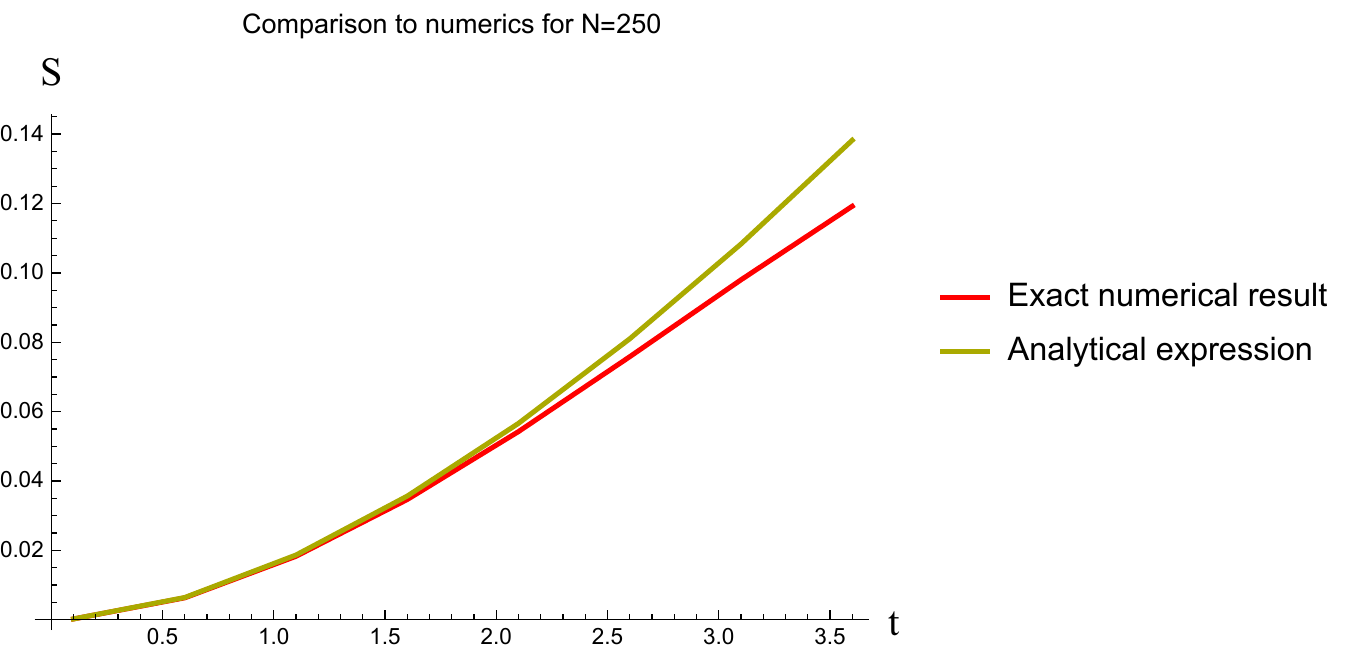}
\caption{Comparison of the analytic expression for $S$ with the one obtained numerically using the matrix (\ref{matrix}) for $N=250$.}
\end{figure}

The situation changes dramatically in the over-critical domain, in which $\epsilon$ becomes imaginary and the system becomes a fast scrambler \cite{scrambling}. 

\subsubsection{The modes contributing to entanglement}
It is clear that formula (\ref{an}) gives an overestimate of the real coefficients, since we ignored the depletion regulating interaction terms. Nevertheless, quite remarkably, even within this approximation  $\kappa_n$ does not grow uncontrollably in time. Indeed, the maximal value that our approximated $\kappa_n$ can attain is
\begin{equation}
\max_t(\,\abs{\kappa_n(t)}^2\,)=\frac{1}{n+1}\left(\frac{n}{n+1}\right)^n\approx\frac{\ee^{-1}}{n+1} \quad \text{for large n}.
\end{equation}
However, since it is $n\abs{\kappa_n}^2$ what enters into the density matrix we find that all modes contribute almost democratically to the entanglement entropy within our approximations.
\subsubsection{Criticality of the ground state}

In the overcritical regime $\epsilon$ becomes imaginary and the homogeneous condensate is unstable. The new ground state is a non-homogeneous condensate, described as a shift-invariant superposition of bright solitons located at different points
on a circle \cite{bright} \footnote{This has interesting implications, like the possibility of preparing Schr\"odinger's  cat states using the gas of bosons. In this paper we shall ignore this subtlety since it does not affect the present discussion}. The ground state for $\lambda \, > \, 1$ is classical in the sense that for large-$N$ it can be well-described by a coherent or a product state for long times.  In this regime, and close to criticality 
($\lambda - 1\, \ll \, 1$), the initially-homogeneous condensate state $\ket{0}$ entangles very efficiently, because of the instability, which provides a Lyapunov exponent, and because 
of the high-density of states near the critical point \cite{scrambling}.   

  We must distinguish this situation from a very high entanglement that takes place  near the critical point 
  when approaching it from the sub-critical regime ($\lambda < 1$).   Although the system can reach a very highly entangled state, the growth 
  of the entanglement in this  regime is slower than in the scrambling regime and takes over a macroscopic time scale 
  $\sim \sqrt{N}$.   
This entanglement  does not occur because the system's ground state is another  classical state, different from $\ket{0}$. Rather, the physical reason for high entanglement is that in the critical regime the ground state is \textit{maximally far from being classical}. In a separate work we will show that in the $N\rightarrow \infty$ limit the $n$ point functions do not factorize for any $n$ and hence the ground state cannot be approximated by ``classical" (e.g., product, coherent, squeezed etc.) states. \\

To conclude this section, the glimpses of black hole type information physics are clearly visible in 
the above analysis of the entropy growth near the quantum-critical point.  From general arguments by Page \cite{page}  
it has been expected that black hole's half life time plays the transitional role in information-processing, 
but the microscopic reason behind this has not been identified clearly in the past.    
 The meaning that quantum-critical portrait of black hole  gives to Page's time is the time of maximal entanglement 
 of black hole constituents, which can be modeled in group-theoretic terms as number of steps required to entangle 
states in the Fock space of gapless qubits of the critical point \cite{infogroup}.   From the above computation of the entanglement entropy evolution, as well as from the previous analysis \cite{nico, scrambling, gold}, it is evident that 
 already the simplest critical condensates capture the properties of development of maxima entanglement over  the power-law 
 time in $N$.

\section{Conclusions and Outlook}

From the recent studies \cite{cesargia, nico, infogroup, scrambling, gold, NV}
 the following is  becoming more and more evident.  First, the peculiar properties of black hole 
information-processing originate from  the fact that black holes represent multi-particle systems 
at quantum critical point.  Secondly, the key properties of black hole information processing are shared by 
other critical systems, such as Bose-Einstein condensates.  This fact opens up an exciting possibility of 
implementing black hole type quantum computing in such systems, both theoretically and experimentally. 

  As it was uncovered by the above studies of the simple prototype models, the generic property is the appearance 
   near the quantum critical point of nearly-gapless weakly-interacting modes that act as qubits for information storage 
   and processing.  In different approaches these modes have been effectively described as Bogoliubov modes of the cold gas of bosons (e.g., gravitons) \cite{cesargia} or as Nambu-Goldstone modes of an isomorphic  sigma-model\cite{gold}.      
   It is evident that seemingly-mysterious properties, such as, e.g., cheap information storage for macroscopically-long time and fast scrambling of information, are determined by the  behavior of these quibits in various near-critical regimes. 
 In particular, in under-critical domain, the qubits store information very cheaply and for a very long time. 
 The ground-state at the critical point is highly entangled, but if we approach the critical point from $\lambda < 1$ 
 regime, the development of the maximal entanglement takes a very long time.  In contrast, in the overcritical regime 
 $\lambda > 1$, Lyapunov exponent ensures the fast scrambling of information \cite{scrambling}. 
  
   All these features, enable us to implement various regimes of black hole quantum computing by manipulating the system parameters externally.  In this paper we achieved the following goals. \\
   
   First,  by coupling the system to an external influence, we 
 designed a simple computation sequence by first encoding information into the Bogoliubov modes and later decoding this information. 
 We showed  how the information is encoded due to interaction of the system 
 with external $c$-modes.  We gave example of  how to  dial $b$-modes to the desired states by performing measurements solely on  $c$-states.  This process of inscription   
 can be viewed as an effective description of information-encodement in a wide range of  critical systems reduced to their bare essentials.   
  For example, for black holes, the role of the $c$-modes can be played by an external scattered 
 radiation, whereas for cold atomic systems, this can be a laser light or another interacting atomic gas. 
 
   We showed, that by tuning parameters of the system, such as $N$ and $\epsilon$, we can make the time evolution of 
   $b$-modes arbitrarily slow, and thus, the information-storage time in these modes -- arbitrarily long. 
   
    Next, using $b$-modes as control qubits for $c$-modes, we can read out the stored information. 
     We gave some examples of control logic gates.  
 
 Finally, by using a new analytic method, to be discussed in more details in \cite{mischa}, we went beyond the Bogoliubov approximation and studied  
 growth of entanglement. We re-confirm previous results \cite{nico, scrambling, gold}, that in sub-critical domain the growth 
 remains slow even beyond the Bogoliubov regime. In the same time, the final level of entanglement 
 is maximal at the critical point.   It is tempting to suggest, following \cite{scrambling, infogroup}, that 
this maximal entanglement of the ground-state near the critical point can shed light at the 
black hole state after its half lifetime (i.e.,  Page's time \cite{page}).  \\
 
  The next obvious task would be to realize, the type of quantum computing sequence considered here,  in an experimental setup.  
      In the view of the progress in quantum gas experiments \cite{coldatoms},  one could envisage 
      designing the systems that can be manipulated near the quantum critical point. 
      Perhaps, this may be attempted by suitably adjusting the setups which allow to design coupled cold gases in one dimension, such as \cite{coupledgases}.  
     Since the type of quantum criticality we considered is based on attractive interaction, it would be interesting to see, if the similar quantum processing can 
      be implemented in systems with stable magnetic droplets observed in \cite{magnet}, which are known to be associated with the attractive 
     interaction between the elementary spin-excitations \cite{magnons} and nucleate at threshold current \cite{magnet1}.

\section*{Acknowledgements}

 We would like to thank Daniel Flassig, Andre Franca, Cesar Gomez and Nico Wintergerst,  for many valuable discussions
 and ongoing collaboration. 
 It is a pleasure to thank Immanuel Bloch and Wilhelm Zwerger for discussions on colds atoms and Andy Kent
 for discussion on magnetic droplet experiments.   
 
The work of G.D. was supported by the Humboldt Foundation under Alexander von Humboldt Professorship, the ERC Advanced Grant ``UV-completion through Bose-Einstein Condensation (Grant No. 339169) and by the DFG cluster of excellence ``Origin and Structure of the Universe", FPA 2009-07908.

\end{document}